\documentclass[%
 reprint,
 amsmath,amssymb,
 aps,
]{revtex4-2}

\usepackage{graphicx}% Include figure files
\usepackage{dcolumn}% Align table columns on decimal point
\usepackage{bm}% bold math
\usepackage{dirtytalk}
\usepackage{siunitx}
\usepackage{braket}
\begin{document}

\preprint{APS/123-QED}

\title{Noncontractible loop states from a partially flat band in a photonic borophene lattice}% Force line breaks with \\

\author{Philip Menz}
\email{philip.menz@uni-muenster.de}
\author{Haissam Hanafi}%
 \email{haissam.hanafi@uni-muenster.de}
\author{J{\"o}rg Imbrock}
\author{Cornelia Denz}
\affiliation{Institute of Applied Physics, University of M{\"u}nster}

\date{\today}% It is always \today, today,
             %  but any date may be explicitly specified

\begin{abstract}
Flat band systems are usually associated to compact localized states (CLSs) resulting from the macroscopic degeneracy of eigenstates at the flat band energy. In case of singular flat bands, these conventional localized flat band states have been found to be incomplete leading to the existence of so-called noncontractible loop states (NLSs) with nontrivial real space topology. Here, we experimentally and analytically demonstrate that a 2D photonic borophene lattice can host NLSs without a CLS counterpart because of a band that is flat only along high-symmetry lines and dispersive along others. We hence dispel the conventional wisdom that NLSs are necessarily linked to robust boundary modes due to a bulk-boundary correspondence. Originating from band touching protected by the band flatness, NLSs play a key role in studying the fundamental physics of flat band systems. Our results truly transform the current understanding of flat bands and could readily be transferred to a 2D material in the form of a planar sheet of boron atoms.
\end{abstract}

%\keywords{Suggested keywords}%Use showkeys class option if keyword
                              %display desired
\maketitle

%\tableofcontents

Flat bands are dispersionless energy bands with vanishing wave group velocity at all momenta in the Brillouin zone. Flat band systems have attracted widespread interest because of their quenched kinetic energy, which maximizes interactions and renders them an ideal playground to study strongly correlated phenomena such as the fractional quantum Hall effect~\cite{Neupert2011}, or superconductivity~\cite{Paul2022}. The presence of a flat band with its macroscopic degeneracy of eigenstates in the spectrum of a periodic lattice implies the existence of so-called compact localized states (CLSs). CLSs are wave functions that are strictly confined to a finite number of lattice sites and remain intact during evolution due to destructive interference. For a lattice of $N$ unit cells, there should be $N$ linearly independent CLSs to span the entire flat band. However, it has been shown that if the flat band has a singular band touching point with a dispersive band, the maximum number of linearly independent CLSs is always less than $N$~\cite{Bergman2008,Rhim2019}. Missing states that complete the basis can be identified as the so-called noncontractible loop states (NLSs). These are states that extend indefinitely along a certain direction, while being localized along the others. They manifest nontrivial real space topology as they span closed loops that wind around the torus representing the lattice with periodic boundary conditions.

Until now, NLSs have always been associated with lattice systems that have flat bands over the entire Brillouin zone. In fact, their presence (absence) is the key signature of the singularity (nonsingularity) of a flat band~\cite{Rhim2021,Hanafi2022,song2022topological}. Here, we break new ground and show that NLS are not only a feature of completely flat bands. In our photonic realization of a chiral borophene lattice, they arise from bands that are partially flat along high-symmetry lines in the Brillouin zone, while dispersive along other directions. Consequently, this kind of flat band does not host any CLSs. Additionally, we reveal that the three NLSs of the chiral borophene lattice are linearly independent and wind around the torus in three topologically distinct directions. This is in stark contrast to previously reported NLSs, which are always only linearly independent in pairs, such as those of the Lieb~\cite{Xia2018}, kagome~\cite{Ma2020}, or super-honeycomb lattice~\cite{Yan2020}. Also, what gives further relevance to our results is the fact that we propose and realize flat band states in the photonic analog of a realistic atomic 2D material, namely a borophene allotrope~\cite{Zope2011a,Yi2017a}.

The theoretical prediction of NLSs is one side of the medal, the other side is a direct experimental observation which proves difficult since NLSs are stable only for an infinite lattice or under periodic boundary conditions. The latter could be realized in atomic borophene in the future, e.g., in the form of fullerene-like or carbon nanotube-like structures. However, the typical lattices considered in flat band experiments are usually finite with open boundaries. This is the case for artificial lattice realizations in form of metamaterials~\cite{Nakata2012}, Bose-Einstein condensates~\cite{Taie2015a}, polariton condensates~\cite{Baboux2016a}, acoustic lattices~\cite{Shen2022}, and photonic waveguide arrays~\cite{Diebel2016b,Travkin2017c,Vicencio2015,Mukherjee2015,hanafi2022localized}, as well as for 2D atomic materials such as the recently realized electronic Lieb and kagome lattices~\cite{Slot2017,Lin2018}. A solution to this problem is to truncate the finite lattice with appropriately tailored boundaries which stabilize the NLSs in the form of \say{line states}. Recently, these line states have been realized in a photonic Lieb lattice~\cite{Xia2018}, and a photonic super-honeycomb lattice~\cite{Yan2020}. Hence, we adopt the versatile technique of femtosecond direct laser writing which enables us to fabricate a finite photonic borophene lattice with tailored open boundaries to stabilize the NLSs. We employ this lattice to experimentally demonstrate the existence of three linearly independent NLSs originating from a partially flat band through their nondiffracting propagation in the form of structured light fields.
%\section{Results and discussion}
\begin{figure*}
\centering\includegraphics[width=\linewidth]{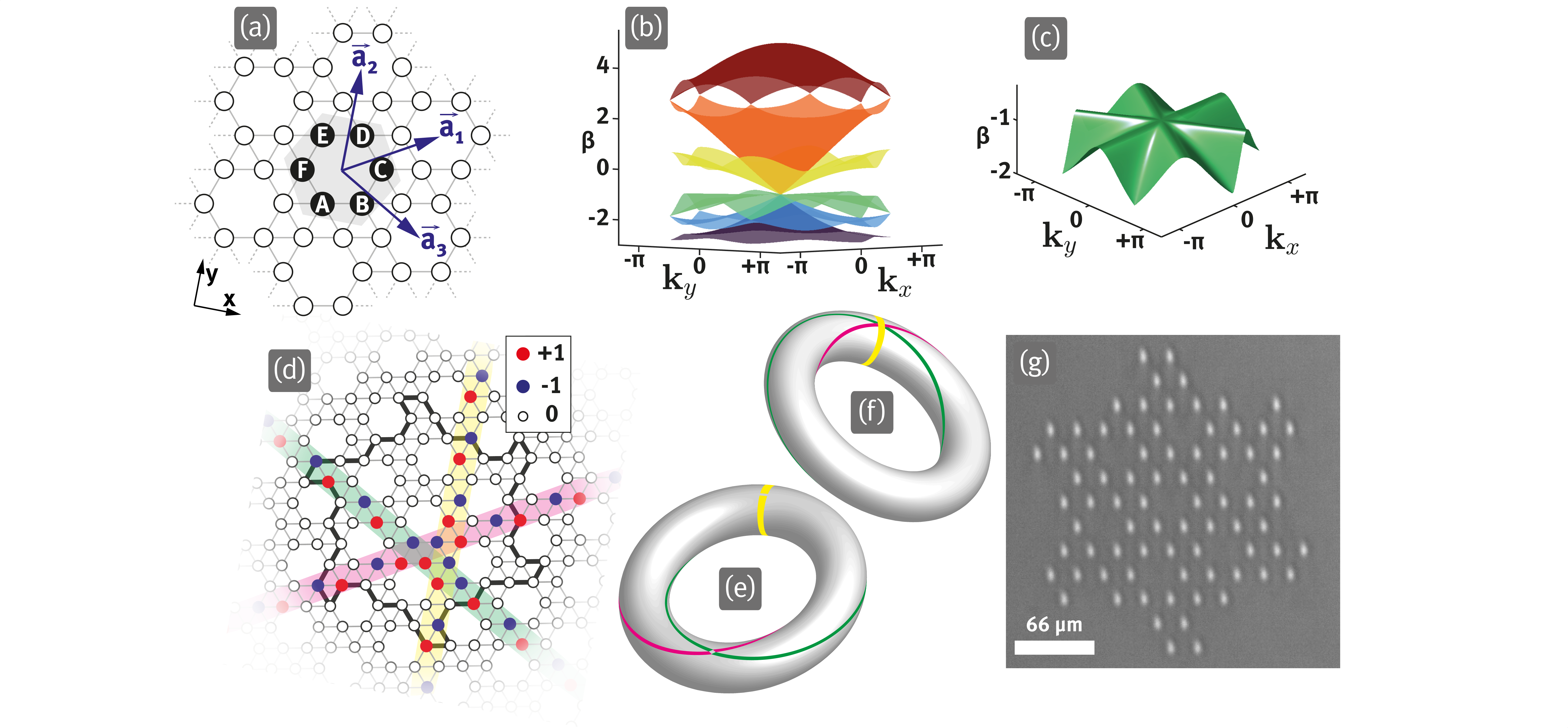} 
\caption{(a) Schematic of the chiral borophene lattice, with the unit cell highlighted in gray and the lattice sites labeled from A to F. (b) Tight-binding band structure in the first Brillouin zone. (c) Partially flat band. (d) The three linearly independent NLSs in an \textit{infinite} lattice, the bold lines indicate the tailored boundaries that stabilize the NLSs in a finite lattice. Lattice sites with zero amplitude are indicated by empty circles, while filled blue and red lattice sites indicate a equal amplitude but opposite phase. (e) and (f) Different perspectives on NLSs winding a torus of periodic boundary conditions. (g) Microscope image of the laser-written chiral borophene lattice in fused silica with tailored boundaries.}
\label{fig:latt_overview}
\end{figure*}

To demonstrate the three linearly independent NLSs arising from a partially flat band, we rely on the borophene lattice shown in Fig.~\ref{fig:latt_overview}(a). In solid states physics, this lattice consisting of a planar sheet of boron atoms is known as the $\chi\textrm{--}h_0$ phase, snub, or chiral borophene~\cite{Zope2011a,Yi2017a} and corresponds to an Archimedean tessellation. This so-called snub hexagonal tiling of the plane is composed of regular triangles and hexagons and exists in two chiral variants~\cite{Grunbaum1977}. The hexagonal unit cell is composed of six lattice sites (A, B, C, D, E, and F) as marked in Fig.~\ref{fig:latt_overview}(a). In a photonic waveguide realization of this lattice, the propagation of light therein is governed by a Schr{\"o}dinger-type paraxial wave equation~\cite{Longhi2009}. Under the tight-binding approximation, a discrete model describing the coupling between the lattice sites can be used. Considering nearest neighbor coupling only, we obtain a $k$-space Hamiltonian
\begin{equation}\label{eq:Ham_snub_boro}
    \hat{H}_\mathbf{k} = t \begin{pmatrix}
0 & 1 & e^{-\text{i}\mathbf{a}_1\mathbf{k}} & e^{-\text{i}\mathbf{a}_2\mathbf{k}} & e^{-\text{i}\mathbf{a}_2\mathbf{k}} & 1 \\
1 & 0 & 1 & e^{-\text{i}\mathbf{a}_2\mathbf{k}} & e^{\text{i}\mathbf{a}_3\mathbf{k}} & e^{\text{i}\mathbf{a}_3\mathbf{k}} \\
e^{\text{i}\mathbf{a}_1\mathbf{k}} & 1 & 0 & 1 & e^{\text{i}\mathbf{a}_3\mathbf{k}} & e^{\text{i}\mathbf{a}_1\mathbf{k}}\\
e^{\text{i}\mathbf{a}_2\mathbf{k}} & e^{\text{i}\mathbf{a}_2\mathbf{k}} & 1 & 0 & 1 & e^{\text{i}\mathbf{a}_1\mathbf{k}}\\
e^{\text{i}\mathbf{a}_2\mathbf{k}} & e^{-\text{i}\mathbf{a}_3\mathbf{k}} & e^{-\text{i}\mathbf{a}_3\mathbf{k}} & 1 & 0 & 1\\
1 & e^{-\text{i}\mathbf{a}_3\mathbf{k}} & e^{-\text{i}\mathbf{a}_1\mathbf{k}} & e^{-\text{i}\mathbf{a}_1\mathbf{k}} & 1 & 0
\end{pmatrix},
\end{equation}
where the lattice vectors are given by $\mathbf{a_1} = d/2(\sqrt{3},1)$, $\mathbf{a_2} = d(0,1)$, and $\mathbf{a_3} = \mathbf{a_1} - \mathbf{a_2}$, $d$ is the lattice constant and $t$ is the coupling strength. Setting $d=t=1$ for simplicity, the eigenvalues of $\hat{H}$ give the spectrum $\beta(\mathbf{k})$ shown in Fig.~\ref{fig:latt_overview}(b). In the case of a photonic lattice, the band structure with its propagation constant $\beta(\mathbf{k})$ represents a diffraction relation that describes the \textit{spatial} evolution dynamics of photonic wave functions in the lattice. In an atomic borophene lattice, this would correspond to an energy spectrum describing the \textit{temporal} evolution of the electronic wave function. Therefore, absence of dispersion in the electronic case traduces to non-diffractive in the photonic one~\cite{Paltoglou2015b}. The band structure is composed of six bands of which five meet at the center of the Brillouin zone, i.e. the $\Gamma$-point, forming a pseudospin-2 conical intersection~\cite{Leykam2016a}. In this work, we focus on the middle band of this higher-order conical intersection, which is partially flat. As shown in Fig.~\ref{fig:latt_overview}(c), the band is perfectly flat at $\beta=-1$ along three high-symmetry lines in the Brillouin zone given by the reciprocal lattice vectors which meet at the singular $\Gamma$-point of fivefold degeneracy. There is an NLS belonging to each of these flat high-symmetry lines resulting from the band touching similarly to previously studied singular flat band lattices like the Lieb~\cite{Xia2016}, or the kagome one~\cite{Ma2020}. However, as the band is not entirely flat, there exist no CLSs. This finding is in stark contrast to the previously described systems. Expressions for the NLSs can be obtained by reducing the 2D to a 1D Hamiltonian. As any 1D flat band has been shown to be nonsingular, it is always possible to find a state that is localized along this dimension and extends indefinitely along the perpendicular direction when going back to the full 2D Hamiltonian~\cite{Rhim2019}. In this way, we can determine the three NLSs that extend unbounded along the lattice vectors $\mathbf{a}_{1,2,3}$, which are shown in Fig.~\ref{fig:latt_overview}(d). These NLSs consist of lattice sites which have the same amplitude but are pairwise out-of-phase satisfying the destructive interference condition that keeps them localized by preventing diffraction via coupling to other waveguides. Remarkably, the three NLSs are linearly independent of each other, which means that it is not possible to obtain any one of them by combining the other two. This can be seen Fig.~\ref{fig:latt_overview}(d) from the fact that the NLSs reside at different lattice sites of the unit cell: A and D for the NLS extended along $\mathbf{a}_3$, B and E for the NLS along $\mathbf{a}_1$, and C and F for the NLS along $\mathbf{a}_2$. This is a new situation compared to NLSs that have been previously reported, which are always only independent in pairs~\cite{Xia2018,Ma2020,Yan2020}. On a torus representing the lattice with periodic boundaries, as can be seen in Figs.~\ref{fig:latt_overview}(e)-(f), the NLSs show a distinct real space topology in their winding around the torus. $\text{NLS}_{\mathbf{a}_2}$ forms a loop along the poloidal direction (yellow), while $\text{NLS}_{\mathbf{a}_1}$ and $\text{NLS}_{\mathbf{a}_3}$ are in the form of Villarceau circles (magenta and green, respectively).

\begin{figure}
\centering\includegraphics[width=\linewidth]{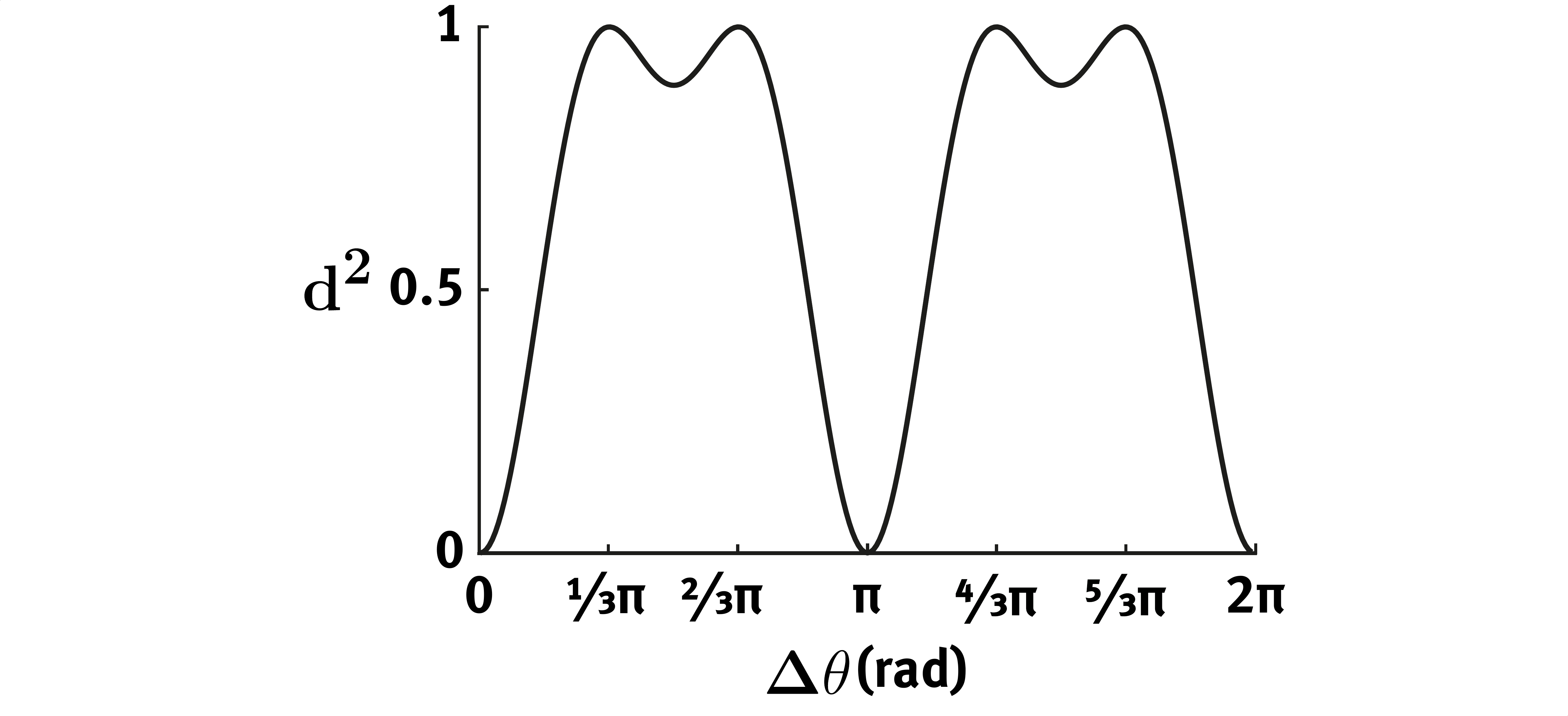} 
\caption{Hilbert-Schmidt quantum distance for two eigenstates of the partially flat band on an arbitrarily close circle around the singular $\Gamma$-point as a function of the angle between them.}\label{fig:Hilb_Schm}
\end{figure}

Further insight into the linear independence of the NLSs can be gained by considering the Hilbert-Schmidt quantum distance of the states of the partially flat band around the band touching singularity. This measure has recently been proposed as a bulk invariant to measure the strength of the singularity at the band crossing point~\cite{Rhim2020}. The Hilbert-Schmidt quantum distance is defined as~\cite{dodonov2000hilbert}
\begin{equation}
    d^2(\psi_1,\psi_2) = 1 - \left | \Braket{ \psi_1 | \psi_2 }  \right |^2,
\end{equation}
where $\psi_1$ and $\psi_2$ are normalized quantum states. In our case, they are eigenstates in momentum space. The quantum distance of two such eigenstates at $\mathbf{k}_1$ and $\mathbf{k}_2$ usually tends to zero for two close momenta $\left | \mathbf{k}_1 - \mathbf{k}_2 \right | \rightarrow 0$. However, around the singular band crossing point of a flat band with a dispersive one, two eigenstates develop a nonzero quantum distance even for really close momenta. In the following we calculate the quantum distance for a pair of eigenstates of our partially flat band on an arbitrarily small circle around the singular $\Gamma$-point with the polar angle $\theta$ at $\mathbf{k}_1 = \epsilon (\cos{\theta_1},\sin{\theta_1})$ and $\mathbf{k}_2 = \epsilon (\cos{\theta_2},\sin{\theta_2})$. For $\epsilon \rightarrow 0$, an eigenstate of the partially flat band of Eq.~\ref{eq:Ham_snub_boro} reads
\begin{equation}
    \ket{\psi(\theta)}=
    \frac{\sqrt{2}}{6}
    \begin{pmatrix}
        -1+2\sin{(2\theta+\pi/6)}\\ 
        1+2\sin{(2\theta-\pi/6)}\\ 
        1+2\cos{(2\theta)}\\ 
        -1-2\cos{(2\theta)}\\ 
        -1-2\sin{(2\theta-\pi/6)}\\ 
        1-2\sin{(2\theta+\pi/6)}
\end{pmatrix}.
\end{equation}
From this, with $\Delta\theta = \theta_1 - \theta_2$, we get $\Braket{\psi_1|\psi_2}=1/3+2/3\cos{(2\Delta\theta)}$, and therefore the quantum distance shown in Fig.~\ref{fig:Hilb_Schm}, which is given by
\begin{equation}
    d^2(\Delta\theta)=\frac{1}{9}\left [ 8 -4\cos{(2\Delta\theta)}-4\cos^2{(2\Delta\theta)} \right ].
\end{equation}
We can see that any two eigenstates are orthogonal to each other at $\Delta\theta = \pi/3 + n\pi$ and $\Delta\theta = 2/3\pi + n\pi$ with $n \in  \mathbb{Z}$. There the quantum distance reaches its maximum value of one. The presence of three orthogonal eigenstates around the degeneracy is a novel situation compared to previously reported singular flat bands where only two of them exist~\cite{Rhim2021}.

To experimentally demonstrate the existence of the three NLSs, we fabricate a sample composed of 78 single-mode waveguides arranged in a chiral borophene lattice geometry. The waveguides extend for \SI{2}{\centi\meter} in the propagation direction $z$ and are induced by femtosecond direct laser writing in fused silica. The fabrication process is carried out with the same parameters reported in~\cite{Hanafi2022}. A microscope image of the front facet of the sample is shown in Fig.~\ref{fig:latt_overview}(g). In a photonic lattice, the waveguide separation distance $\Lambda$ allows to control the coupling strength. To ensure sufficient coupling during propagation in the sample while minimizing unwanted next-nearest neighbor coupling, we choose a value of $\Lambda = \SI{22}{\micro\meter}$. To stabilize the NLSs in a finite open boundary lattice, we choose the appropriate tailored termination shown in Fig.~\ref{fig:latt_overview}(g). An indirect demonstration of the NLSs in the form of so-called robust boundary modes (RBMs) is not possible in our case since no CLSs exist and RBMs and CLSs are related by a bulk-boundary correspondence~\cite{Ma2020}.

To observe the nondiffracting propagation of the NLSs in the photonic lattice, we shape probe light fields in the form of the NLSs by generating Gaussian spots of alternating phase with an SLM. The spots are arranged to match the positions of the waveguides corresponding to the NLSs.
\begin{figure*}
\centering\includegraphics[width=\linewidth]{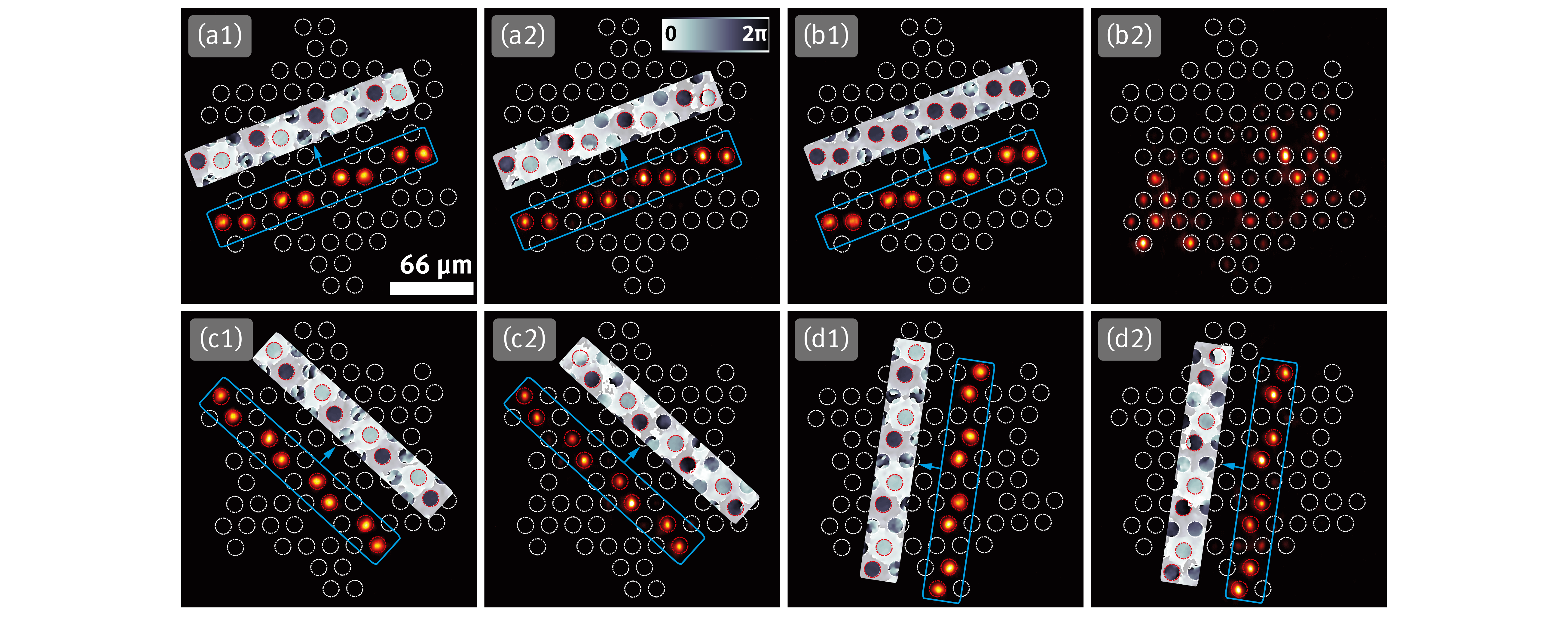} 
\caption{Experimental demonstration of diffractionless propagation of the NLSs in the chiral borophene photonic lattice. (a1) Intensity of the $\text{NLS}_{\mathbf{a}_1}$ input light field with the phase shown in the inset. (a2) Same as (a1) but after propagation through the photonic lattice. (b1)-(b2) Same as (a1)-(a2), but for a diffracting state which has equal phase at every lattice site. (c1)-(c2) Same as (a1)-(a2), but for $\text{NLS}_{\mathbf{a}_3}$. (c1)-(c2) Same as (a1)-(a2), but for $\text{NLS}_{\mathbf{a}_2}$.}.\label{fig:exp_res}
\end{figure*}
The first example light field corresponding to $\text{NLS}_{\mathbf{a}_1}$ is shown in Fig.~\ref{fig:exp_res}(a1). As can be seen in the output recorded after propagation through the lattice in Fig.~\ref{fig:exp_res}(a2), the NLS stays completely localized during propagation in the photonic lattice as the destructive interference prevents it from diffracting. The out-of-phase relation is perfectly preserved as shown in the insets obtained by a digital holographic interference measurement~\cite{Kreis1986}. For comparison, the results for an input of in-phase spots are shown in Fig.~\ref{fig:exp_res}(b1)-(b2). In this case, because the input light field is not a flat band one but is composed of modes from dispersive bands, there is considerable diffraction during propagation by coupling to previously unexcited waveguides. We then proceed to experimentally demonstrate the two remaining NLSs. The input and output light fields of $\text{NLS}_{\mathbf{a}_3}$ are shown in Fig.~\ref{fig:exp_res}(c1)-(c2), respectively. Again, after propagation through the lattice, the intensity stays completely localized in the initially excited waveguides, as the out-of-phase relation is perfectly preserved. The results for $\text{NLS}_{\mathbf{a}_2}$ are shown in Fig.~\ref{fig:exp_res}(d1)-(d2). They thoroughly confirm the expectation of diffractionless propagation inherent to a flat band state. With this observations we are able to confirm the existence of NLSs originating from a partially flat band without CLSs. An additional remarkable novelty of the NLSs of the chiral borophene lattice is that three of them are linearly independent of each other. This can be seen from the fact that they reside in different waveguides of the unit cell of the photonic lattice. The three observed linearly independent NLSs extend along different directions in the lattice. In a periodic boundary picture, they exhibit a distinct real space topological winding around the torus.

%\section{Conclusion}
In conclusion, the results presented here reshape the current understanding of wave localization in flat band systems by showing that NLSs are not exclusively a feature of completely flat bands. The demonstrated NLSs originate from a band that is only partially flat along three high-symmetry lines in the Brillouin zone that meet at a fivefold degenerate conical intersection. Due to the non-flatness of the whole band, there exist no CLSs and consequently no robust boundary modes can be found. Our work opens an avenue to study unconventional wave localization in lattices that are not usually considered as flat band ones, e.g., in the context of topological insulators~\cite{biesenthal2022fractal,pyrialakos2022bimorphic}. An important and new feature of our NLSs is the linear independence of three of them, meaning that they exhibit a distinct nontrivial real space topological winding. The realization of NLSs in a photonic version of a borophene allotrope brings exciting new possibilities for flat band physics in atomic 2D materials beyond artificial lattice platforms. Our results suggest that the newly found flat band states could be observed in an electronic setting. They are of relevance for the field of electronic flat band systems and highly correlated states that are considered powerful guidelines for the search of room-temperature superconductivity~\cite{torma2022superconductivity}.
\bigskip

P.M. and H.H. contributed equally to this work.

 %\bibliography{NLS_paper}% Produces the bibliography via BibTeX.

%%%%%%%%%%%%%%%%%%%%%%%%%%%%%%%%%%%%%%%%%
%%%%%%%%%%%%%%%%%%%%%%%%%%%%%%%%%%%%%%%%%

%apsrev4-2.bst 2019-01-14 (MD) hand-edited version of apsrev4-1.bst
%Control: key (0)
%Control: author (8) initials jnrlst
%Control: editor formatted (1) identically to author
%Control: production of article title (0) allowed
%Control: page (0) single
%Control: year (1) truncated
%Control: production of eprint (0) enabled
%

%%%%%%%%%%%%%%%%%%%%%%%%%%%%%%%%%%%%%%%%%
%%%%%%%%%%%%%%%%%%%%%%%%%%%%%%%%%%%%%%%%%

\end{document}